\documentstyle[12pt,amsbsy]{article}
\newcommand{\beq}{\begin{equation}}
\newcommand{\eeq}{\end{equation}}

\newcommand{\beqn}{\begin{eqnarray}}
\newcommand{\eeqn}{\end{eqnarray}}

\newcommand{\p}{\mbox{${\bf p}$}}
\newcommand{\q}{\mbox{${\bf q}$}}
\newcommand{\r}{\mbox{${\bf r}$}}
\newcommand{\n}{\mbox{${\bf n}$}}
\newcommand{\bk}{\mbox{${\bf k}$}}
\newcommand{\bv}{\mbox{${\bf v}$}}
\newcommand{\s}{\mbox{${\bf s}$}}
\newcommand{\bS}{\mbox{${\bf S}$}}
\newcommand{\si}{\mbox{${\boldsymbol\sigma}$}}
\newcommand{\Si}{\mbox{${\boldsymbol\Sigma}$}}
\newcommand{\vrho}{\mbox{${\boldsymbol\rho}$}}
\newcommand{\vzeta}{\mbox{${\boldsymbol\zeta}$}}
\newcommand{\vom}{\mbox{${\boldsymbol\omega}$}}
\newcommand{\bu}{\mbox{${\bf u}$}}
\newcommand{\E}{\mbox{${\bf E}$}}
\newcommand{\B}{\mbox{${\bf B}$}}

\newcommand{\f}{\mbox{${\bf f}$}}
\newcommand{\fib}{\mbox{${\boldsymbol\phi}$}}
\newcommand{\va}{\mbox{${\bf a}$}}
\newcommand{\ep}{\mbox{${\epsilon}$}}

\begin{document}

\begin{titlepage}

\vspace{1cm}

\begin{center}
{\Large \bf  Budker Institute of Nuclear Physics}
\end{center}

\vspace{0.5cm}

\begin{flushright}
BINP 97-81\\
October 1997
\end{flushright}

\vspace{1cm}

\begin{center}
{\large \bf Equations of motion of spinning relativistic particle\\
in external fields}
\end{center}

\begin{center}
I.B. Khriplovich\footnote{khriplovich@inp.nsk.su} and
A.A. Pomeransky\footnote{pomeransky@vxinpz.inp.nsk.su}
\end{center}
\begin{center}
Budker Institute of Nuclear Physics\\
630090 Novosibirsk, Russia
\end{center}

\bigskip

\begin{abstract}
We consider the motion of a spinning relativistic particle in
external electromagnetic and gravitational fields, to first order
in the external field, but to an arbitrary order in spin.
The correct account for the spin influence on the particle
trajectory is obtained with the noncovariant description of spin.
Concrete calculations are performed up to second order in spin
included. A simple derivation is presented for the gravitational
spin-orbit and spin-spin interactions of a relativistic particle.
We discuss the gravimagnetic moment (GM), a specific spin effect
in general relativity. It is demonstrated that for the Kerr black
hole the gravimagnetic ratio, i.e., the coefficient at the GM,
equals to unity (as well as for the charged Kerr hole the
gyromagnetic ratio equals to two). The equations of motion obtained
for relativistic spinning particle in external gravitational field
differ essentially from the Papapetrou equations.

\end{abstract}

\vspace{3.5cm}

\end{titlepage}

\section{Introduction}

The problem of motion of a particle with internal angular momentum
(spin) in an external field consists of two parts: the description
of the spin precession and the account for the spin influence on
the trajectory of motion. To lowest nonvanishing order in $c^{-2}$
the complete solution for the case of an external electromagnetic
field was given more than 70 years ago \cite{tho}. The gyroscope
precession in a centrally symmetric gravitational field had been
considered to the same approximation even earlier \cite{des}. Then,
essentially later the spin precession was investigated in the case
of the gravitational spin-spin interaction \cite{sch}. The fully
relativistic problem of the spin precession in an external
electromagnetic
field was also solved more than 70 years ago \cite{fr}, and then in
a more convenient formalism, using the covariant vector of spin, in
Ref. \cite{bmt}.

The situation with the second part of the problem, which refers to
the spin influence on the trajectory, is different. Covariant
equations of motion for a relativistic spinning particle in an
electromagnetic field were written in the same paper \cite{fr}, and
for the case of a gravitational field in Ref. \cite{pa}. These
equations have been discussed repeatedly from various points of
view in numerous papers [7--15].
The problem of the spin influence on the trajectory of a particle in
external fields is not of a purely theoretical interest only. It
attracts attention being related to the description of the motion
of ultrarelativistic particles in accelerators \cite{dk} (see also
recent review \cite{hei}). Then, there are macroscopic objects for
which internal rotation influences their motion in an external
gravitational field. We mean Kerr black holes. This problem is of
importance in particular for the calculation of the gravitational
radiation of binary stars. In this connection it was considered in
Refs. [18--21]. However, when turning to these
calculations, we found out \cite{khp} that the equations of motion
with the account for spin to the lowest nonvanishing order in
$c^{-2}$, used in these papers, even in the simpler case of an
external field lead to results which differ from the well-known
gravitational spin-orbit interaction. The problem is essentially
related to the correct definition of the center-of-mass coordinate.
It turned out moreover that the Papapetrou equations \cite{pa} as
well do not reproduce in the same $c^{-2}$ approximation the result
for the gravitational spin-orbit interaction ascending to Ref.
\cite{des}. This discrepancy was pointed out long ago in Ref.
\cite{bar}, however its explanation suggested there does not look
satisfactory (see Ref. \cite{khp}).

In the present work we obtain equations of motion of a
relativistic particle with the noncovariant description of spin.
They agree with well-known limiting cases. Though for external
electromagnetic field such equations in the linear in spin
approximation have been obtained previously \cite{dk} (see also
Ref. \cite{hei}), we
would like to start with comments related to this approximation
in electrodynamics.

\bigskip

\section{Covariant and noncovariant equations of motion \\
         of spinning particle in electromagnetic field }

The interaction of spin with external electromagnetic field
is described, up to terms on the order of $c^{-2}$ included,
by the well-known Hamiltonian
\begin{equation}\label{th}
H=-\,\frac{eg}{2m}\s\B+\,
\frac{e(g-1)}{2m^2}\s[\p \times \E] \,.
\end{equation}
Let us emphasize that the structure of the second, Thomas,
term in this expression not only is firmly established
theoretically, but has been also confirmed with high accuracy
experimentally, at any rate in atomic physics. The force
corresponding to the Hamiltonian (\ref{th}) equals
\begin{equation}\label{thac}
\f_m=\,\,\frac{eg}{2m}\s\B,_m+\,\frac{e(g-1)}{2m}
\,\left(\frac{d}{dt}[\E \times \s\,]_m-\,
\s[\bv \times \E,_m\,]\right);
\end{equation}
here and below coma with index denotes a partial derivative.

Let us try to construct a covariant equation of motion
accounting for spin, which would reproduce in the same
approximation the force (\ref{thac}). A covariant correction
$f^{\mu}$ to the Lorentz force $eF^{\mu\nu}u_{\nu}$, linear in
the tensor of spin $S_{\mu\nu}$ and in the gradient of the
tensor of electromagnetic field $F_{\mu\nu,\lambda}\,$, may
depend also on the 4-velocity $u^\mu$. Since
$u^{\mu}u_{\mu}=1$, this correction should satisfy the
condition $u_{\mu}f^{\mu}=0$. From the mentioned tensors one
can construct only two independent structures which satisfy
the last condition. The first one,
$$\eta^{\mu\kappa}F_{\nu\lambda,\kappa}S^{\nu\lambda}\,-\,
F_{\lambda\nu,\kappa}u^{\kappa}S^{\lambda\nu}u^{\mu},$$
reduces in the approximation of interest to
$$2\s(\B,_m-\,[\bv \times \E,_m]),$$
and the second,
$$u^{\lambda}F_{\lambda\nu,\kappa}u^{\kappa}S^{\nu\mu},$$
reduces to
$$\frac{d}{dt}[\s \times \E]\,_m\,.$$
(Let us note that the structures with the product
$F_{\nu\kappa,\lambda}S^{\kappa\lambda}$ reduce to the two
presented expressions, due to the Maxwell equations and the
antisymmetry of $S_{\kappa\lambda}$.) Obviously, no linear
combination of the two presented structures can reproduce
the correct expression (\ref{thac}) for the spin-dependent
force. In a somewhat less general way it was demonstrated
in Ref. \cite{khp}. It was pointed out in the same paper,
that the coordinate entering the covariant equation does
not coincide with the usual one. Therefore, to obtain the
correct $c^{-2}$ approximation to the covariant equation
of motion one needs to perform an additional redefinition
of coordinate:
\beq\label{rx}
\r \rightarrow \r - \frac{1}{2m}\bv \times \s.
\eeq
In the case of spin $1/2$ this redefinition is closely
related to the Foldy-Wouthuysen transformation \cite{fw}.
The generalization of this substitution to the case of
arbitrary velocities was recently suggested in Ref.
\cite{hei}.

Meanwhile, the correct equations of motion in electromagnetic
field including spin to first order are known for sufficiently
long time \cite{dk}. Let us recall that the initial physical
definition of spin refers to the proper frame of the particle.
This is the 3-dimensional vector $\s$ (or 3-dimensional
antisymmetric tensor) of the internal angular momentum defined
in this frame. The covariant vector of spin $S_{\mu}$ (or the
covariant antisymmetric tensor $S_{\mu\nu}$) are obtained from
them just by the Lorentz transformation. By the way, in this
approach the constraints
$u^{\mu}S_{\mu}=0$, $u^{\mu}S_{\mu\nu}=0$ are valid identically.
The precession frequency for spin $\s$ at an arbitrary velocity
is well-known (see, for instance, Ref. \cite{blp}):
$$ {\bf \Omega}=\,\frac{e}{2m}\,\left\{(g-2)\,\left[\B
-\,\frac{\gamma}{\gamma+1}\,\bv(\bv\B)\, -\,\bv \times \E
\right]\,\right.$$
\beq\label{Om}
\left.+2\,\left[\frac{1}{\gamma}\,\B
-\,\frac{1}{\gamma+1}\,\bv \times \E\right]\right\},
\eeq
where $\gamma =\,1/\sqrt{1-v^2}$. Naturally, the corresponding
interaction Lagrangian (the Lagrangian description is here
somewhat more convenient than the Hamiltonian one) equals
\[ L_{1s} = {\bf \Omega}\s=\,\frac{e}{2m}\,\s\,
\left\{(g-2)\,\left[\B
-\,\frac{\gamma}{\gamma+1}\,\bv(\bv\B)\, -\,\bv \times \E
\right]\,\right. \]
\beq\label{lse}
\left.+2\,\left[\frac{1}{\gamma}\,\B
-\,\frac{1}{\gamma+1}\,\bv \times \E\right]\right\}.
\eeq
The force, we are looking for, is given by the usual relation
\beq\label{fs}
\f = (\nabla - \frac{d}{dt} \nabla_{\bv})L,
\eeq
and the equation of motion for spin in general form is
\beq\label{dsc}
\dot{\s} = - \{L,\s\},
\eeq
where $\{...\;,...\}$ denotes the Poisson brackets, or
\beq\label{dsq}
\dot{\s} = - i[L,\s]
\eeq
in a quantum problem.

In the conclusion of this section, let us discuss the following
question. In fact, it is far from being obvious how meaningful
are at all the discussed spin corrections to the equations of
motion of elementary particles, say, electron or proton.
According to the well-known argument by Bohr (see Ref.
\cite{pau}), an additional Lorentz force due to the finite
size of the wave packet of a charged particle and to the
uncertainty relation, exceeds the corresponding component
of the Stern-Gerlach force. Meanwhile, it was proposed long
ago to separate by polarizations a beam of charged
particles in a storage ring through the spin interaction
with external fields \cite{ros}. Though this proposal is
being discussed rather actively (see review \cite{hei}),
it is not clear up to now whether it is feasible in principle.
In this connection we would like to note that, as distinct
from the Lorentz force $e(\E+\bv \times \B)$, the component
of the spin force (\ref{fs}) directed along the velocity,
grows with energy, its asymptotic behaviour being
$$\f \rightarrow
\gamma\bv\,\left(\frac{\partial}{\partial t} +
\,\bv \nabla\right)\,\frac{e}{m}\,\s \left[\B-\,\bv \times \E
+\,\frac{g-2}{2}\,\bv(\bv\B)\right].$$
This growth of the spin force with energy gives at least
in principle the possibility to circumvent the limitation
due to the uncertainty relation.

\bigskip

\section{Equations of motion of spinning particle\\
         in electromagnetic field. General formalism. \\
         Effects linear in spin}

In this section we point out a general approach to the
derivation of the equations of motion in external
electromagnetic field to an arbitrary order in spin.
We will reproduce here in passing the known result
(\ref{Om}).

We derive the Lagrangian of the spin interaction
with an external field from the scattering amplitude
\beq\label{ampl}
- e J^{\mu}A_{\mu}
\eeq
of a particle with spin $s$ off a vector potential
$A_{\mu}$. Due to the arguments presented in the end
of the previous section, the account for the effects
nonlinear in spin, which we are first of all
interested in, may have physical meaning only in the
classical limit $s\gg 1$. Just this approximation
will be mainly used below.

The matrix element $J_{\mu}$ of the electromagnetic
current operator between states with momenta $k$ and
$k^\prime$ can be written (under $P$ and $T$ invariance)
as follows (see, for instance, Refs. \cite{khri,kms}):
\beq\label{cur}
J_{\mu}=\frac{1}{2\ep}\,\bar{\psi}(k^\prime)
\left\{p_{\mu}F_e\right.
\left. +\, \Sigma_{\mu\nu}q^{\nu}\,F_m\right\}\,\psi(k).
\eeq
Here $\;p_{\mu}=(k^\prime+k)_{\mu},
\;\; q_{\mu}=(k^\prime-k)_{\mu}$.

The wave function of a particle with an arbitrary spin
$\psi$ can be written (see, for instance, Ref.
\cite{blp}, \S 31) as
\begin{equation}\label{ps}
            \psi={1 \over \sqrt 2} \left( \begin{array}{c} \xi\\
                                         \eta\\
                        \end{array}
                 \right).
\end{equation}
Both spinors,
\[ \xi=\{ \xi^{{\alpha}_1\,{\alpha}_2\,\,..\,\,
{\alpha}_p\,}_{\dot{\beta}_1
\,\dot{\beta}_2\,..\,\,\dot{\beta}_q}\} \]
and
\[ \eta=\{ \eta_{\dot{\alpha}_1\,\dot{\alpha}_2
\,\,..\,\,\dot{\alpha}_p\,}^{{\beta}_1\,{\beta}_2\,..\,\,
{\beta}_q}\},
\]
are symmetric in the dotted and undotted indices separately,
and
\[ p+q=2s. \]
For a particle of half-integer spin one can choose
\[ p=s+\,\frac{1}{2}\,,\;\;\;\;q=s-\,\frac{1}{2}\,\,. \]
In the case of integer spin it is convenient to use
\[ p=q=s. \]
The spinors $\,\xi\,$ and $\,\eta\,$ are chosen in such a way
that under reflection they go over into each other (up to a
phase). At $p \neq q$ they are different objects which
belong to different representations of the Lorentz group.
If $p=q$, these two spinors coincide. Nevertheless, we will
use the same expression (\ref{ps}) for the wave function of
any spin, i.e., we will also introduce formally the object
$\,\eta\,$ for an integer spin, keeping in mind that it is
expressed in terms of $\,\xi\,$. This will allow us to
perform calculations in the same way for the integer and
half-integer spins.

In the rest frame both $\xi$ and $\eta$ coincide with a
nonrelativistic spinor $\xi_0$, which is symmetric in all
indices; in this frame there is no difference between dotted
and undotted indices. The spinors $\xi$ and $\eta$ are
obtained from $\xi_0$ through the Lorentz transformation:
\beq\label{lt}
\xi=\exp\{\Si\fib/2\} \xi_0\,;\;\;\;\;\;\;\;
\eta=\exp\{-\Si\fib/2\} \xi_0\,.
\eeq
Here the vector $\fib$ is directed along the velocity,
$\;\;\tanh\phi=v$;
\[  \Si \,=\,\sum_{i=1}^{p} \si_i\,-\,
\sum_{i=p+1}^{p+q} \si_i\,, \]
and $\si_i$ acts on the $i$th index of the spinor $\,\xi_0\,$
as follows:
\begin{equation}\label{aa}
\si_i\,\xi_0
=(\si_i)_{\alpha_i\beta_i}\, (\xi_0)_{....\beta_i...}\;.
\end{equation}
In the Lorentz transformation (\ref{lt}) for $\xi$, after
the action of the operator $\Si$ on $\xi_0$ the first $p$
indices are identified with the upper undotted indices and
the next $q$ indices are identified with the lower dotted
indices. The inverse situation takes place for $\eta$.

Let us note that in an external field the components of
velocity $\bv$ (and together with them the components of
$\fib$) do not commute, generally speaking. However, to the
approximation linear in external field, we are interested
in, one can neglect this noncommutativity which is itself
proportional to the field. Moreover, we are mainly interested
in the classical limit of the final result
where such commutators are inessential
being proportional to an extra power of $\hbar$. Therefore,
$\bv$ and $\fib$ will be treated as usual numerical parameters.

Then,
\[ \bar{\psi} = \psi^\dagger \gamma_0 =
             \psi^\dagger  \left(
                            \begin{array}{rr}
                                0 & I \\
                                I & 0\\
                             \end{array}
                       \right);
      \]
here $I$ is the sum of unit $2\times 2$ matrices acting on
all indices of the spinors $\,\xi\,$ and $\,\eta\,$. The
components of the matrix $\Sigma_{\mu\nu}=-\Sigma_{\nu\mu}$
are:
\begin{equation}\label{0n}
      \Sigma_{0n}= \left(
                            \begin{array}{rr}
                             -\Sigma_n    & 0 \\
                                 0 & \Sigma_n\\
                             \end{array}
                       \right);
      \end{equation}
\begin{equation}\label{mn}
      \Sigma_{mn}=\,-\,2i\ep_{mnk}\left(
                            \begin{array}{rr}
                                s_k & 0  \\
                                0   & s_k\\
                             \end{array}
                       \right);
      \end{equation}
\[ \s =\,\frac{1}{2}\sum_{i=1}^{2s} \si_i.  \]

The scalar operators $F_{e,m}$ depend on two invariants,
$t=q^2$ and $\tau=(S^{\mu}q_{\mu})^2$. The covariant vector
of spin $S_{\mu}$ is defined, e.g., for the state with
momentum $k_{\mu}$, and is obtained through the Lorentz
transformation from the spin vector $(0,\s)$ in the rest
frame:
\beq\label{Ss}
S^{\mu}=(S_0,\bS),\;\;S_0=\frac{(\s\bk)}{m},\;\;
 \bS=\s+\frac{\bk(\bk\s)}{m(\ep+m)}.
\eeq
In the expansion in the electric multipoles
\[ F_e(t,\tau)=\sum_{n=0}^{N_e}f_{e,2n}(t)\tau^n \]
the highest power $N_e$ equals obviously to $s$ and
$s-1/2$ for integer and half-integer spin, respectively.
In the magnetic multipole expansion
\[ F_ m(t,\tau)=\sum_{n=0}^{N_m}f_{m,2n}(t)\tau^n \]
the highest power $N_m$ constitutes $s-1$ and $s-1/2$ for
integer and half-integer spin. It can be easily seen that
$$f_{e,0}(0)=1,\;\;\;\;f_{m,0}(0)=\frac{g}{2}.$$

Let us note at last that we have chosen the noncovariant
normalization for the amplitude (\ref{ampl}), being
interested in the Lagrangian referring to the world time
$t$, but not to the proper time $\tau$.

Let us reproduce now in this approach the well-known result
(\ref{lse}) for the case of a constant external field. We
will start with the terms proportional to $g$-factor. The
corresponding term in the scattering amplitude can be
written as
\[\frac{eg}{4\ep}\;\xi_0^{\prime \dagger}
\left\{[\exp\{\Si\fib/2\}(\s\B)
\exp\{-\Si\fib/2\}+\exp\{-\Si\fib/2\}(\s\B)\exp\{\Si\fib/2\}]
\right.\]
\beq\label{lseg}
\left.+\,{i \over 2}
\,[\exp\{\Si\fib/2\}(\Si\E)\exp\{-\Si\fib/2\}
-\exp\{-\Si\fib/2\}(\Si\E)\exp\{\Si\fib/2\}]\right\}\xi_0.
\eeq
It is essential that in the case considered, that of the
constant external field, one may put
$\;{\bf k}^\prime={\bf k},\;{\bf v}^\prime
={\bf v},\;\fib^\prime=\fib\,$, since $\;{\bf q}
={\bf k}^\prime-{\bf k}\,$ corresponds to the field gradient.

The further calculations use the known identity
\[ \exp\{\hat A\} \hat B \exp\{-\hat A\}
= \hat B\,+\,\frac{1}{1!}\,[\hat A,\,\hat B]\,
+\,\frac{1}{2!}\,\left[\hat A,\,[\hat A,\,\hat B]\right]+...\; ,\]
as well as the relations
\beq\label{ide1}
 [\Sigma_i,\,\Sigma_j]=4i\ep_{ijk}s_k\, ,\;\;\;
   [\Sigma_i,\,s_j]=i\ep_{ijk}\Sigma_k\, ;
\eeq
\beq\label{ide2}
\cosh \phi = \gamma,\;\;\; \sinh \phi = v\,\gamma.
\eeq
After sufficiently simple algebraic transformations
expression (\ref{lseg}) reduces to
\beq\label{lseg1}
\frac{eg}{2m}\,\s \left[\B
-\,\frac{\gamma}{\gamma+1}\,\bv(\bv\B)\, -\,\bv \times \E
\right].
\eeq

Let us go over now to the contribution of the convection term
\beq\label{conv}
-\,\frac{e}{2\ep}\,\bar{\psi}(k^\prime)\psi(k)\,p^{\mu}A_{\mu}.
\eeq
We rewrite the product of exponents in the expression
\beq\label{pro}
\bar{\psi}(k^\prime)\psi(k)= {1 \over 2}\xi_0^{\prime \dagger}
[\exp\{\Si\fib^\prime/2\} \exp\{-\Si\fib/2\}
+\exp\{-\Si\fib^\prime/2\}\exp\{\Si\fib/2\}]\xi_0
\eeq
as
\[ \exp\{\Si\fib^\prime/2\} \exp\{-\Si\fib/2\} \]
\beq\label{prod}
=\prod_p\exp\{\si\fib^\prime/2\} \exp\{-\si\fib/2\}
\prod_q\exp\{-\si\fib^\prime/2\} \exp\{\si\fib/2\}.
\eeq
Let us consider a typical factor in this formula:
\[ \exp\{\si\fib^\prime/2\} \exp\{-\si\fib/2\}
= \cosh(\phi^\prime/2)\cosh(\phi/2)
-(\n^\prime \n)\sinh(\phi^\prime/2)\sinh(\phi/2)\]
\beq\label{exp}
+\si\left[\n^\prime\sinh(\phi^\prime/2)\cosh(\phi/2)
-\n\cosh(\phi^\prime/2)\sinh(\phi/2)\right]
\eeq
\[-\,i\,(\si[\n^\prime \times \n])\sinh(\phi^\prime/2)\sinh(\phi/2);\]
here $\;\n^\prime=\bv^\prime/v^\prime,\;\n=\bv/v\,$. Since we
are interested in gradients only as long as they enter together
with spin, in the first term
$\;\cosh(\phi^\prime/2)\cosh(\phi/2)
-(\n^\prime \n)\sinh(\phi^\prime/2)\sinh(\phi/2)\;$ we put
$\;\phi^\prime=\phi/2,\;\;\n^\prime = \n\, ,$ after which this
term turns to unity. Then, we are discussing  the interaction
linear in spin, so that the product (\ref{prod}) reduces to
\[ 1 + \Si\left[\n^\prime\sinh(\phi^\prime/2)\cosh(\phi/2)
-\n\cosh(\phi^\prime/2)\sinh(\phi/2)\right]
-\,2i\,(\s[\n^\prime \times \n])\sinh(\phi^\prime/2)\sinh(\phi/2).\]
When substituted into formula (\ref{pro}), the terms
proportional to $\Si$ cancel. Restricting then to the terms
linear in $\q$, we reduce the spin-dependent part of
(\ref{conv}) to
\[ -\,e\,\frac{p^{\mu}}{2\ep}\,\frac{i(\s[\bk \times\q])}
{m(\ep+m)}\,A_{\mu}. \]
Let us note further that since $\;p^{\mu}q_{\mu}=0\,$, the
identity takes place
\beq\label{ide}
p^{\mu}q_{\alpha}A_{\mu}=p^{\mu}(q_{\alpha}A_{\mu}-q_{\mu}A_{\alpha})
=p^{\mu}iF_{\alpha\mu}.
\eeq
Now we can put $\;p_{\mu}\rightarrow 2 m u_{\mu}\,,$
where $\,u_{\mu}\,$ is the 4-velocity. In result we arrive at
the following expression:
\beq\label{lse1}
-\,\frac{e}{2m}\,\s \left[2\,\left(1\,-\,{1\over \gamma}\right)\B
-\,\frac{2\gamma}{\gamma+1}\,\bv(\bv\B)\,
-\,\frac{2\gamma}{\gamma+1}\,\bv \times \E
\right].
\eeq
Expressions (\ref{lseg1}), (\ref{lse1}) give in sum formula
(\ref{lse}). Thus, we have reproduced the known result for the
interaction linear in spin, starting from the relativistic
wave equation for an arbitrary spin.

Below we will use repeatedly identities of the type (\ref{ide}).
In the classical language such a transformation corresponds to
omitting in a Lagrangian (or adding to it) a total time derivative.
Indeed,
\[ u^{\mu}q_{\mu}\rightarrow u^{\mu}\partial_{\mu}
=\,\gamma\,\left({\partial \over \partial t}\,+\,\bv \nabla\right)\,
=\,\gamma\,{d \over d t}. \]

\bigskip

\section{Equations of motion of spinning particle\\
         in electromagnetic field. Effects quadratic in spin}

Let us go over now to the interaction of second order in spin.
The "bare", explicit quadrupole interaction present in
expressions (\ref{ampl}), (\ref{cur}) is
\beq\label{qu}
-\,e\,{p^{\mu} \over 2\ep}\,f_{e,2}(S^{\alpha}q_{\alpha})^2
\,A_{\mu}.
\eeq
Using identity (\ref{ide}) and relations (\ref{Ss}), we reduce
it to
\[ -\,e\,f_{e,2}\,\gamma\,\left\{(\bv\s)\left[{\partial \over
\partial t}\,+\,{\gamma \over \gamma+1}\,(\bv \nabla)\right]\,+
\,{1 \over \gamma}(\s\nabla)\right\}\,\left[(\s\E)-\,
{\gamma \over \gamma+1}\,(\s\bv)(\bv\E)\,
+\,(\s[\bv\times\B])\right]. \]
Neglecting the total time derivative
$\partial/\partial t\,+\,\bv \nabla$, we arrive at expression
\beq\label{qu1}
L_{2s}=\,-\,e\,f_{e,2}\,\left[(\s\nabla)\,-\,
{\gamma \over \gamma+1}\, (\bv\s)(\bv \nabla)\right]\,\left[(\s\E)\,
-\,{\gamma \over \gamma+1}\,(\s\bv)(\bv\E)\,
+\,(\s[\bv\times\B])\right].
\eeq
Using the Maxwell equations and adding a total derivative in
$t$, one can demonstrate that this expression has such a
structure that the tensor $s_i s_j$ in it can be rewritten in
the irreducible form:
$s_i s_j\rightarrow s_i s_j-(1/3)\delta_{ij}\s^2$. Now it is
clear from the nonrelativistic limit of formula (\ref{qu1})
that it describes indeed the interaction with external field of
the quadrupole moment
\beq\label{qua}
Q_{ij}\,=\,-\,2\,e\,f_{e,2}\,(3\,s_i s_j-\,\delta_{ij}\s^2);
\;\;\;Q\,=\,Q_{zz}\vert_{s_z=s}=\,-\,2\,e\,f_{e,2}\,s(2s-1).
\eeq
In the asymptotics, as $\gamma\rightarrow \infty$, the
interaction (\ref{qu1}) tends to a constant
\beq\label{quas}
L_{2s}=\,-\,e\,f_{e,2}\,[(\s\nabla)\,-\,(\bv\s)(\bv \nabla)]\,
[(\s\E)\,-\,(\s\bv)(\bv\E)\,+\,(\s[\bv\times\B])].
\eeq

It is well-known that even in the absence of the bare
quadrupole term, i.e., at $f_{e,2}=0$, a quadrupole interaction
arises in the nonrelativistic limit due to the convection and
magnetic terms in interaction (\ref{ampl}). The value of this,
induced quadrupole moment at an arbitrary spin of the particle
was obtained in Ref. \cite{kms}$\,$\footnote{The authors of
this paper overlooked a misprint in the answer for the induced
quadrupole moment, in result its value presented in Ref.
\cite{kms} is two times smaller than the correct one}:
\beqn\label{qua1}
Q_1=\,-\,e\,(g-1)\,\left({\hbar \over mc}\right)^2\,
\left\{\begin{array}{ll}
                               s, &  \mbox{integer spin,}\\
                               s-1/2,&\mbox{half-integer spin.}\\
                   \end{array}
             \right.
\eeqn
We have singled out explicitly in this formula the Planck
constant $\hbar$ to demonstrate that the induced quadrupole
moment $Q_1$ vanishes in the classical limit
$\,\hbar\rightarrow 0,\;
s\rightarrow \infty,\; \hbar s\rightarrow$ const. Therefore,
the interaction of second order in spin proportional to $Q_1$
does not influence in fact equations of motion of a classical
particle (though it plays a role in atomic spectroscopy
\cite{kms}).

Still, the convection and magnetic terms in expression
(\ref{ampl}) induce an interaction of second order in spin
which has a classical limit and is therefore of interest for
our problem. It is convenient here to start with the convection
current interaction. Let us come back to formula (\ref{exp}).
Again we put in it
\[ \cosh(\phi^\prime/2)\cosh(\phi/2)
-(\n^\prime \n)\sinh(\phi^\prime/2)\sinh(\phi/2)\,=\,1. \]
And in other terms, linear in $\si$, we keep the first power of
$\q\rightarrow - i\hbar \nabla$ only, in the hope that in the final
answer, in the product (\ref{prod}), $\hbar$ will enter in the
combination $\hbar s\rightarrow$ const. Nevertheless, these
terms by themselves are small as compared to unity, so that in
the classical limit expression (\ref{exp}) can be rewritten as
\[ \exp\left\{
\si\left[\n^\prime\sinh(\phi^\prime/2)\cosh(\phi/2)
-\n\cosh(\phi^\prime/2)\sinh(\phi/2)\right]
-\,i\,(\si[\n^\prime \times \n])\sinh^2(\phi/2)\right\}. \]
It can be easily seen that in the product (\ref{prod}) the
operators $\si$, entering with
\[ \n^\prime\sinh(\phi^\prime/2)\cosh(\phi/2)
-\n\cosh(\phi^\prime/2)\sinh(\phi/2)\,,\]
combine in the resulting exponent into the operator $\Si$
which vanishes in the classical limit. In this limit only
those operators $\si$ survive which enter with
$\;[\n^\prime \times \n]\sinh^2(\phi/2);\;$
they combine into $2\s$. Thus, in the classical limit the
product (\ref{prod}) reduces, with the account for the second
identity (\ref{ide2}) to
\beq\label{Prod}
\exp\left\{ {\gamma \over \gamma+1}\,\left(\s\,[\bv\times
\nabla]\right)\right\}.
\eeq

Let us note that that the action of the operator (\ref{Prod}) on
any function of coordinates, whether it is a vector potential or
field strength, reduces to the shift of its argument:
\[ \r \rightarrow \r + {1 \over m}\,{\gamma \over \gamma+1}\,
\s \times \bv\,. \]
Curiously, just this substitution is suggested in Ref. \cite{hei}
for the transition from the covariant equations linear in spin
to noncovariant ones. Its particular case in the $c^{-2}$
approximation is formula (\ref{rx}).

Now, taking into account the second term in the expansion of the
exponential function (\ref{Prod}) and using again the identity
(\ref{ide}), we easily obtain the following expression for the
quadratic in spin interaction arising from the convection
current:
\beq\label{s2c}
-\,{e \over 2m^2}\,{\gamma \over \gamma+1}\,\left(\s\,[\bv\times
\nabla]\right)\left[\left(1-\,{1 \over \gamma}\right)(\s\B)\,-\,
{\gamma \over \gamma+1}\,(\s\bv)(\bv\B)\,-\,
{\gamma \over \gamma+1}\left(\s\,[\bv\times \E]\right)\right].
\eeq

Let us go over now to the contribution into the discussed
effect due to the magnetic moment. The term in the scattering
amplitude due to the magnetic moment, we are interested in,
can be conveniently written now as
\[\frac{eg}{4\ep}\;\xi_0^{\prime \dagger}
\left\{[\exp\{\Si\fib^\prime/2\}\exp\{-\Si\fib/2\}
\exp\{\Si\fib/2\}(\s\B)\exp\{-\Si\fib/2\}\right. \]
\[ +\exp\{-\Si\fib^\prime/2\}\exp\{\Si\fib/2\}
\exp\{-\Si\fib/2\}(\s\B)\exp\{\Si\fib/2\}] \]
\beq\label{lseg2}
+\,{i \over 2}\,[\exp\{\Si\fib^\prime/2\}\exp\{-\Si\fib/2\}
\exp\{\Si\fib/2\}(\Si\E)\exp\{-\Si\fib/2\}
\eeq
\[ \left. -\exp\{-\Si\fib^\prime/2\}\exp\{\Si\fib/2\}
\exp\{-\Si\fib/2\}(\Si\E)\exp\{\Si\fib/2\}]\right\}\xi_0. \]
Using in this case the first term in the expansion of the
exponential function (\ref{Prod}), we arrive at the following
expression for the contribution proportional to the magnetic
moment:
\beq\label{s2g}
\,{eg\over 2m^2}\,{\gamma \over \gamma+1}\,\left(\s\,[\bv\times
\nabla]\right)\left[(\s\B)\,-\,
{\gamma \over \gamma+1}\,(\s\bv)(\bv\B)\,
-\,\left(\s\,[\bv\times \E]\right)\right].
\eeq
The total result for the induced interaction, quadratic in spin,
is
\[ L_{2s}^i=\,{e \over 2m^2}\,{\gamma \over \gamma+1}
\,\left(\s\,[\bv\times \nabla]\right)
\left[\left(g-1+\,{1 \over \gamma}\right)(\s\B)\,-\,
(g-1)\,{\gamma \over \gamma+1}\,(\s\bv)(\bv\B)\,\right. \]
\beq\label{qu2}
\left. -\,\left(g-{\gamma \over \gamma+1}\right)
\left(\s\,[\bv\times \E]\right)\right].
\eeq

Let us note that in the nonrelativistic limit the induced
interaction with magnetic field tends to zero as $v/c$, and
that with electric field as $(v/c)^2$. Moreover, the part of
interaction (\ref{qu2}) which is not related to $g$-factor,
is reducible in spin; in other words, the structure
$\,s_i s_j\,$ in it cannot be rewritten as an irreducible tensor
$\,s_i s_j - (1/3)\delta_{ij}\s^2\,$. The interaction (\ref{qu2})
is not at all a quadrupole one. However, its asymptotic
behaviour at $\,\gamma\rightarrow \infty\,$ is of interest. In
this limit
\beq\label{quasi}
L_{2s}^i=\,{e \over 2m^2}\,(g-1)\,\left(\s\,[\bv\times \nabla]\right)
\,[(\s\B)\,-\,(\s\bv)(\bv\B)\,-\,
\left(\s\,[\bv\times \E]\right)].
\eeq
Surprisingly, the asymptotical formulae (\ref{quas}) and
(\ref{quasi}) coincide up to a factor and a total time derivative.
To prove this fact, it is convenient to introduce three
orthogonal unit vectors
\[ \bv;\;\;\vrho\,=\,{[\bv\times \s] \over |[\bv\times \s]|};\;\;
   \vzeta\,=\,[\bv\times \vrho]. \]
Using the completeness of this basis and the equation $\;\dot{\E}=
[\nabla\times \B]\,,$ and neglecting as well a total derivative
in $\,t\,$, one can easily check that
\[ [(\s\nabla)\,-\,(\bv\s)(\bv \nabla)]\,
[(\s\E)\,-\,(\s\bv)(\bv\E)\,+\,(\s[\bv\times\B])] \]
\[ =\,[\bv\times \s]^2\,(\vzeta\nabla)[(\vzeta\E)\,+\,(\vrho\B)]\;, \]
coincides indeed with
\[ \left(\s\,[\bv\times \nabla]\right)
\,[(\s\B)\,-\,(\s\bv)(\bv\B)\,
-\,\left(\s\,[\bv\times \E]\right)] \]
\[ =\,-\,[\bv\times \s]^2\,(\vrho\nabla)
\left[\left(\vrho[\bv\times \B]\right)\,+\,(\vrho\E)\right]. \]
Thus, there is a special value of the bare quadrupole moment
(\ref{qua})
\beq\label{quab}
Q\,=\,-\,2(g-1)\,{es^2 \over m^2}\,,\;\;\;\mbox{or}\;\;
f_{e,2}\,=\,(g-1)\,{1 \over 2m^2}
\eeq
(let us recall that we consider now a classical situation, when
$\,s\gg 1\,$), at which the total interaction, quadratic in
spin, $\,L_{2s}+L_{2s}^i\,$, falls down asymptotically with
energy.

The situation resembles that which takes place for the
interaction linear in spin. It is well-known (see, for instance,
\cite{kh,wei,fpt}) that there is a special value of $g$-factor,
$\,g=2\,$, at which the interaction linear in spin decreases
at $\;\gamma\rightarrow \infty\,$. It follows immediately from
formula (\ref{lse}) for the first-order Lagrangian. Thus,
putting additionally $\,g=2\,$, we obtain
\beq\label{quab1}
Q\,=\,-\,2\,{es^2 \over m^2}\,,\;\;\;\mbox{or}\;\;
f_{e,2}\,=\,{1 \over 2m^2}\,.
\eeq

Let us note that the choice $\,g=2\,$ for the bare magnetic
moment is a necessary (but insufficient) condition of
renormalizability in quantum electrodynamics \cite{kh,wei,fpt}.
It is satisfied not only for electron, but also for the charged
vector boson in the renormalizable electroweak theory.

In some respect, however, the situation with the singled out
values (\ref{quab}), (\ref{quab1}) of the quadrupole moment
differs from the situation with $g$-factor. Conditions
(\ref{quab}), (\ref{quab1}), as distinct from the condition
$\,g=2\,$, are not universal, they take place only for large
spins, $\;s\gg 1\,$, in other words, they refer only to
classical objects with internal angular momentum. In
particular, for the charged vector boson of the renormalizable
electroweak theory the bare quadrupole interaction is absent
at all, $\;f_{e,2}=0\,$. The quadrupole moment of this particle
is (in our language) of the induced nature, it is given by
formula (\ref{qua1}) at $\,s=1\,$ É $\,g=2\,$.

\bigskip

\section{Spin precession in gravitational field}

In this section we present a simple and general derivation of
the equations of the spin precession in a gravitational field.
This approach not only allows one to reproduce and easily
generalize known results for spin effects. Pointed out here
remarkable analogy between gravitational and electromagnetic
fields allows one to transcribe easily the results of two
previous sections for the case of an external gravitational
field.

It follows from the angular momentum conservation in flat
space-time taken together with the principle of equivalence that
the spin 4-vector $S^\mu$ is parallel transported along the
particle world-line. The parallel transport of a vector
along a geodesic $x^\mu(\tau)$ means that its covariant
derivative vanishes:
\beq\label{par}
\frac{DS^\mu}{D\tau}=\,0\,.
\eeq
Let us go over to the tetrad formalism natural for the
description of spin. In virtue of relation (\ref{par}) the
equation for the tetrad components of spin
$S^a=\,S^\mu e^a_{\mu}$ is
\begin{equation}\label{pars}
\frac{DS^a}{D\tau}=\,\frac{dS^a}{d\tau}=\,S^\mu e^a_{\mu;\nu}u^\nu=
\,\eta^{ab}\gamma_{bcd}u^d S^c\,.
\end{equation}
Here
\beq\label{rota}
\gamma_{abc}=\,e_{a\mu;\nu}e^\mu_{b}e^\nu_{c}=\,-\gamma_{bac}
\eeq
are the Ricci rotation coefficients \cite{ll}. Certainly, the
equation for the tetrad 4-velocity components is exactly the
same:
\begin{equation}\label{paru}
\frac{du^a}{d\tau}=\,\eta^{ab}\gamma_{bcd}u^d u^c\,.
\end{equation}
The meaning of Eqs. (\ref{pars}), (\ref{paru}) is clear: the
tetrad components of both vectors vary in the same way, due to
the rotation of the local Lorentz vierbein only.

In the exactly same way, the 4-dimensional spin and velocity of
a charged
particle with the gyromagnetic ratio $g=2$ precess with the same
angular velocity in an external electromagnetic field, in virtue
of the Bargman, Michel, Telegdi equation \cite{bmt,blp} (at
$g=2$) and the Lorentz equation:
\[ \frac{dS_a}{dt}=\,\frac{e}{m}F_{ab}S^b;\;\;\;\;\;
\frac{du_a}{dt}=\,\frac{e}{m}F_{ab}u^{b}.\]
Therefore, the evident correspondence takes place:
\beq\label{corcov}
\frac{e}{m}F_{ab} \longleftrightarrow \gamma_{abc}u^c.
\eeq
It allows one to obtain the precession frequency $\vom$ of the
3-dimensional spin vector $\s$ in external gravitational field from
expression (\ref{Om}) by means of simple substitution
\beq\label{cornon}
\frac{e}{m}B_i \longrightarrow
-\,\frac{1}{2}\epsilon_{ikl}\gamma_{klc}u^c;
\;\;\; \frac{e}{m}E_i \longrightarrow \gamma_{0ic}u^c.
\eeq
This frequency equals
\begin{equation}\label{og1}
\omega_i=\,-\epsilon_{ikl}\left(\frac{1}{2}\gamma_{klc}+
\,\frac{u^k}{u^0+1}\gamma_{0lc}\right)\,\frac{u^c}{u^0_w}\,.
\end{equation}
The overall factor $1/u^0_w$ in this expression is related to the
transition in the lhs of Eq. (\ref{pars}) to the differentiation in
the world time $t$:
$$\frac{d}{d\tau}=\,\frac{dt}{d\tau}\,\frac{d}{dt}
=\,u^0_w\,\frac{d}{dt}.$$
The quantity $u^0_w$ is supplied with the subscript $w$ to
emphasize that this is a world component of 4-velocity, but not a
tetrad one. All other indices in expression (\ref{og1}) are tetrad
ones, $c=0,1,2,3;\;\;i,k,l=1,2,3$.
The corresponding spin-dependent correction to the Lagrangian is
\begin{equation}\label{sg1}
L_{1sg}=\,\s \vom\,.
\end{equation}

As an illustration of formulae (\ref{og1}), (\ref{sg1}), let us
apply them to the cases of spin-orbit and spin-spin interactions.
We will restrict, as it is usual in the problem discussed, to the
linear approximation in the gravitational field. However, in our
approach, as distinct from the standard ones, both problems admit
an elementary solution at an arbitrary particle velocity.

Tetrads $e_{a\mu}$ are related to the metrics as follows:
$$e_{a\mu}e_{b\nu}\eta^{ab}=\,g_{\mu\nu}.$$
To linear approximation one can put
$g_{\mu\nu}=\,\eta_{\mu\nu}+\,h_{\mu\nu}$ and do not distinguish
between the tetrad and world indices in $e_{a\mu}$.
The ambiguity in the choice of tetrads will be fixed by choosing
the symmetric gauge $e_{\mu\nu}=\,e_{\nu\mu}\,.$ Then
$$e_{\mu\nu}=\,\eta_{\mu\nu}+\,\frac{1}{2}h_{\mu\nu}\,.$$
Using expression (\ref{rota}) for the Ricci coefficients, we
find to linear approximation
\begin{equation}\label{gam}
\gamma_{abc}=\,\frac{1}{2}(h_{bc,a}-\,h_{ac,b})\,.
\end{equation}

Let us start with the spin-orbit interaction. In the centrally
symmetric field created by a mass $M$, the metrics is
\begin{equation}\label{lin}
h_{00}=\,-\frac{2kM}{r};\;\;\; h_{mn}=\,-\frac{2kM}{r}\delta_{mn}.
\end{equation}
Here nonvanishing Ricci coeffeicients are
\begin{equation}\label{ric}
\gamma_{ijk}=\,\frac{kM}{r^3}(\delta_{jk}r_i-\,\delta_{ik}r_j)\,,
\;\;\;\;\;\gamma_{0i0}=\,-\frac{kM}{r^3}r_i\,.
\end{equation}
Their substitution into formula (\ref{og1}) gives the following
expression for the precession frequency:
\begin{equation}\label{so}
{\vom_{ls}}=\,\frac{2\gamma+\,1}{\gamma+\,1}\,\frac{kM}{r^3}\,
\bv \times \r \,.
\end{equation}
In the limit of low velocities, $\gamma \rightarrow 1$, the
answer goes over into the classical result \cite{des}.

We go over now to the spin-spin interaction. Let the spin of
the central body be $\s _0$. Linear in $\s _0$ components of
metrics, which are responsible for the spin-spin interaction,
are:
$$h_{0i}=\,2k\,\frac{[\s _0 \times \r]_i}{r^3}\,.$$
Here nonvanishing Ricci coefficients equal
\begin{equation}\label{ri1}
\gamma_{ij0}=\,k(\nabla_i\frac{[\s _0\times\r]_j}{r^3}-\,
\nabla_j\frac{[\s _0\times \r]_i}{r^3})\,,\;\;\;\;\;
\gamma_{0ij}=\,-k\nabla_i\frac{[\s _0 \times \r]_j}{r^3}\,.
\end{equation}
The frequency of the spin-spin precession is
$$ \vom_{ss}=\,-k\,\left(2-\frac{1}{\gamma}\right)
(\s _0\nabla)\nabla\frac{1}{r}$$
\begin{equation}\label{ss}
+\,k\,\frac{\gamma}{\gamma+\,1}
\left[\bv (\s _0\nabla) -\,\s _0(\bv \nabla)
+\,(\bv \s _0)\nabla\right]\,(\bv \nabla)\,\frac{1}{r}\,.
\end{equation}
In the limit of low velocities this formula also goes over into
the corresponding classical result \cite{sch}.

In the conclusion of this section let us note that in the case of
an external gravitational field there is no covariant
expression for the force linear in the particle spin. In other
words, the deviation from geodesics of the trajectory of a
spinning particle is not described by the Riemann tensor. In this
case there is a unique possible covariant structure, up to a
factor (in Ref. \cite{pa} it equals $-\,1/2m$):
$R_{\mu\nu ab}u^{\nu}S^{ab}$. As it was mentioned already in
Introduction, that covariant description (as distinct from our
our formulae (\ref{og1}), (\ref{sg1})) contradicts the classical
results in the limit of low velocities.

\bigskip

\section{Equations of motion of spinning particle\\
         in gravitational field. General approach}

Equations of motion in an external gravitational field to any
order in spin are constructed analogously to the equations of
motion in the case of electromagnetic field.

We write down the scattering amplitude in a weak external
gravitational field $h_{\mu\nu}$ as
\beq\label{gampl}
-\,{1 \over 2}T_{\mu\nu}h^{\mu\nu}.
\eeq
The matrix element $T_{\mu\nu}$ of the energy-momentum tensor
between the states of momenta $k$ and $k^\prime$ can be written
as
\[ T_{\mu\nu}=\frac{1}{4\ep}\,\bar{\psi}(k^\prime)
\left\{p_{\mu}p_{\nu}\,F_1
 +\,{1 \over 2}\,(p_{\mu}\Sigma_{\nu\lambda}
+\,p_{\nu}\Sigma_{\mu\lambda})\,q^{\lambda}\,F_2 \right. \]
\beq\label{tens}
\left. +\,(\eta_{\mu\nu}q^2\,-\,q_\mu q_\nu)\,F_3
+\,[S_\mu S_\nu q^2 -\,(S_\mu q_\nu+\,S_\nu q_\mu)(Sq)\,+\,
\eta_{\mu\nu}(Sq)^2]\,F_4 \right\}\,\psi(k).
\eeq
The scalar operators $F_i$ in this expression are also expanded
in powers of $\tau=(Sq)^2$:
\[ F_i(t,\tau)=\sum_{n=0}^{N_i}f_{i,2n}(t)\tau^n. \]
It is easy to convince oneself that the total number of invariant
form factors $f_{i,2n}$ is $4s+2$ and $4s+1$ for integer and
half-integer spin, respectively. The independence of the four
tensor structures entering expression (\ref{tens}) is obvious. As
to the completeness of the expansion, it can be checked, for
instance, by demonstrating that the calculation of the total
number of invariant form factors in the annihilation channel leads
to the same result as the above one.

In the generally covariant form the structure
$(\eta_{\mu\nu}q^2\,-\,q_\mu q_\nu)\,h^{\mu\nu}$ corresponds to
the scalar curvature $R$, and
$[S_\mu S_\nu q^2 -\,(S_\mu q_\nu+\,S_\nu q_\mu)(Sq)\,+\,
\eta_{\mu\nu}(Sq)^2]\,h^{\mu\nu}$ corresponds to the product
$R_{\mu\nu} S^\mu S^\nu$, where $R_{\mu\nu}$ is the Ricci tensor.
Since we are interested in the equations of motion in a
sourceless field, the corresponding terms in the expansion
(\ref{tens}) will be omitted.

As well as in electrodynamics the charge conservation dictates
the condition $\,f_{e,0}(0)=1\,$, here the energy
conservation leads to $\,f_{1,0}(0)=1\,$. As to the term in the
amplitude (\ref{gampl}) which contains $\,f_{2,0}\,$, it is
convenient to rewrite it otherwise, using the analogy
(\ref{corcov}) with electromagnetic field. Substituting into the
corresponding electromagnetic term
\[ i\,\frac{eg}{8\ep}\,\bar{\psi}(k^\prime)\,\Sigma^{ab}
F_{ab}\,\psi(k) \]
$g=2,\;(e/m)F_{ab}=f_{ab}=\gamma_{abc}u^c\,$, we arrive at the
following contribution to the Lagrangian of gravitational
interaction:
\beq\label{gamp2}
i\,\frac{1}{4u^0_w}\,\bar{\psi}(k^\prime)\,\Sigma^{ab}f_{ab}\,
\psi(k);
\eeq
here, as usual, $\;u_w^0=\,\ep/m$. Using for
$\,\gamma_{abc}\;$ the linear approximation (\ref{gam}), one can
check easily that the expression (\ref{gamp2}) corresponds indeed
to the discussed contribution to the amplitude at the condition
$\;f_{2,0}=1$. Thus, in gravity the value of one more form
factor at zero momentum transfer $t$ is fixed. It corresponds to
the conservation of
angular momentum. This circumstance was pointed out many years
ago in Ref. \cite{kook}.

Let us come back to the convection term in formula (\ref{gampl}).
As it was the case in electrodynamics, here when going over to
spinors in the rest frame, the term of first order in spin is
written as
\beq\label{gcon}
-\,{p^{\mu}p^{\nu} \over 8 \ep}\,{1 \over m}\,{u^0 \over u^0+1}
\left(\s\,[\bv\times \nabla]\right)\,h_{\mu\nu}.
\eeq
Using relations (\ref{ide}), (\ref{gam}), we obtain
\[ p^\mu \nabla_k h_{\mu\nu}\rightarrow
-\,p^\mu (-\partial_k h_{\mu\nu}+\,\partial_\mu h_{k\nu})
\rightarrow -\,2 p^a \gamma_{ak\nu}.\]
Thus the expression (\ref{gcon}) is presented through the Ricci
coefficients:
\beq\label{gcon1}
{1 \over u^0_w}\,{u^0 \over u^0+1}\,\ep^{mnk} s^m v^n
u^a u^c \gamma_{akc}.
\eeq
As can be easily checked, the expressions (\ref{gamp2}) and
(\ref{gcon1}) in sum reproduce the Lagrangian (\ref{sg1}).

\bigskip

\section{Equations of motion of spinning particle\\
         in gravitational field. Second order in spin}

Let us investigate now the effects of second order in spin in
the equations of motion in a gravitational field. In the case  
of a binary star these effects are of the same order of magnitude
as the spin-spin interaction at comparable spins of components of
the system \cite{khp}. The influence of the latter on the
characteristics of gravitational radiation becomes noticeable for
a system of two extreme black holes \cite{kww}. Correspondingly,
the spin effects of second order in the equations of motion get
appreciable if at least one component of a binary is an extreme
black hole \cite{khp}. Therefore, the investigation of these
effects is not of a purely theoretical interest only. In
principle they can be observed with the gravitational wave
detectors under construction at present.

An obvious source of the second order spin effects is the term
\beq\label{agm}
L_{2sg}=\,-\,f_{1,2}\,{1 \over 8\ep}\,p^\mu p^\nu (Sq)^2 h_{\mu\nu}
\eeq
in the amplitude (\ref{gampl}). Due to the relation
\[ p^\mu p^\nu q_\alpha q_\beta h_{\mu\nu}=\,
 p^\mu p^\nu ( q_\alpha q_\beta h_{\mu\nu}+
 \,q_\mu q_\nu h_{\alpha\beta}-\,
 q_\alpha q_\nu h_{\mu\beta}-\,q_\beta q_\mu h_{\nu\alpha})
 \rightarrow 2 p^\mu p^\nu R_{\mu\alpha\nu\beta}, \]
the Lagrangian (\ref{agm}) is rewritten via the Riemann tensor:
\beq\label{agmr}
L_{2sg}=\,-\,{\kappa  \over 2\ep }\,u^a S^b u^c S^d R_{abcd}.
\eeq
Instead of $\,f_{1,2}\,$, we have introduced a dimensionless
parameter $\kappa$:
\[ f_{1,2}=\,{\kappa \over 2m^2}. \]

Then it is convenient to use the Petrov representation for the
components of the Riemann tesor (see Ref. \cite{ll}):
\[ E_{kl}=\,R_{0k0l},\;\;E_{kl}=\,E_{lk};\;\;\;
   C_{kl}=\,\frac{1}{4}\epsilon_{kmn}\epsilon_{lrs}R_{mnrs},\;\;
   C_{kl}=C_{lk}; \]
\beq\label{pe1}
B_{kl}=\,\frac{1}{2}\epsilon_{lrs}R_{0krs},\;\;B_{kk}=\,0.
\eeq
We restrict to the case of a sourceless gravitational field.
Then, at $R_{ab}=0$, additional simplifications take place:
\beq\label{pe2}
C_{kl}=\,-\,E_{kl},\;\;\;B_{kl}=\,B_{lk},
\;\;\;E_{kk}=\,C_{kk}=\,0.
\eeq
Finally, we arrive at the following interaction Lagrangian
quadratic in spin:
\[ L_{2sg}=\,-\,{\kappa \over 2\ep}\,
\left[(2\bu^2+1)\,E_{kl}-\,2\,\left(2
-\,{1 \over u_0+\,1}\right) u_k u_m E_{lm}
+\,\delta_{kl}u_m u_n E_{mn}\right. \]
\beq\label{sg2}
+\,{1 \over (u_0+\,1)^2}\,u_k u_l u_m u_n E_{mn}
\eeq
\[ \left. -\,2\,u_0\,\epsilon_{kmn}u_m B_{nl}
+\,{2 \over u_0+\,1}
\,u_k u_m \epsilon_{lrn} u_r B_{mn}\right]\,
(s_k s_l - {1 \over 3}\,\delta_{kl} \s ^2). \]
To avoid misunderstanding, we note that all three-dimensional
indices in this equation (and next two ones) are in fact 
contravariant.

As well as in electrodynamics, here along with the "bare"
interaction (\ref{sg2}), there is an induced one quadratic in
spin. Its explicit form can be obtained most easily by putting
in electromagnetic formula (\ref{qu2}) $g=2$ and by making the
substitution (\ref{cornon}). We take into account also the
correspondence
\[ q_i \gamma_{abc}u^c\,=\,(q_i \gamma_{abc}
-\,q_c \gamma_{abi})\,u^c\,\rightarrow
\,i\,(\partial_i \gamma_{abc}
-\,\partial_c \gamma_{abi})\,u^c\,\rightarrow
\,i\,R_{abci}u^c. \]
Finally, using relations (\ref{pe1}), (\ref{pe2}), we obtain the
following result for the induced interaction:
\[ L_{2sg}^i\,=\,{1 \over 2\ep}\,
\left\{\left(2\bu^2-\,{u^0-1 \over u^0+\,1}\right)\,E_{kl}
-\,2\,\left[2
-\,{1 \over u_0+\,1}-\,{1 \over (u_0+\,1)^2}\right]\,
 u_k u_m E_{lm}\right. \]
\beq\label{sg2i}
+\,\left[1-\,{1 \over (u_0+\,1)^2}\right]\,
\delta_{kl}u_m u_n E_{mn}
+\,{1 \over (u_0+\,1)^2}\,u_k u_l u_m u_n E_{mn}
\eeq
\[ \left. -\,2\,\left(u_0-\,{1 \over u_0+\,1}\right)
\epsilon_{kmn}u_m B_{nl}
+\,{2 \over u_0+\,1}
\,u_k u_m \epsilon_{lrn} u_r B_{mn}\right\}\,s_k s_l. \]

As well as in the electromagnetic case, the induced interaction
here tends to zero in the nonrelativistic limit  $\,\sim v/c\,$,
and the spin factor in it, $\,s_k s_l\,$, is not an irreducible
tensor.

The asymptotical behaviour of $L_{2sg}$ and $L_{2sg}^i$ is the
same: both Lagrangians increase linearly with energy.
However, in this case also the coefficient in the
"bare" interaction can be chosen in such a way, $\,\kappa=1\,$,
that the total Lagrangian of second order in spin decreases
(as well as the analogous interaction in electrodynamics)
when energy tends to infinity. At $\,\kappa=1\,$
\[ L_{2sg}+\,L_{2sg}^i\,=\,-\,{1 \over \ep (u^0+\,1)}\,
\left(u^0\,E_{kl}-\,{1 \over u_0+\,1}\,u_k u_m E_{lm}\right. \]
\beq\label{}
\left. +\,{1 \over 2(u_0+\,1)}\,\delta_{kl}u_m u_n E_{mn}
+\,\epsilon_{kmn}u_m B_{nl} \right)\,s_k s_l.
\eeq

\bigskip

\section{Gravimagnetic moment. Multipoles of black holes}

There is a deep analogy between the linear in spin interaction
of the magnetic moment with the electromagnetic field
\beq\label{Le}
{\cal L} _{em}=-\frac{eg}{4m}F_{ab}S^{ab}\,
\eeq
and the "bare" gravitational Lagrangian (\ref{agmr}) quadratic
in spin \cite{kh}. (It is more convenient here to write the
gravitational Lagrangian, like $\,{\cal L} _{em}\,$, for the
proper time $\,\tau\,$, and not for the world time $\,t\,$, i.e.,
to multiply expression (\ref{agmr}) by $\,\ep/m\,$.) This
analogy is based on the following observation. It is well-known
that the canonical momentum $\,i\partial_\mu\,$ enters
relativistic wave equations for a particle in electromagnetic and
gravitational external fields via the combination
\[ \Pi_\mu = \,i\partial_\mu - eA_\mu -
{1 \over 2}\Sigma^{ab}\gamma_{ab\mu}. \]
It follows from the structure of the commutator (or Poisson
brackets in the classical limit)
\[ [\Pi_\mu,\Pi_\nu]=\,-\,i\,(eF_{\mu\nu}
-\,{1 \over 2}\Sigma^{ab}R_{ab\mu\nu}) \]
that in a sense $\,-\,{1 \over 2}
\Sigma^{ab}R_{ab\mu\nu}\,$ plays the same role in gravity as
$\,eF_{\mu\nu}\,$ in electromagnetism. It is quite natural then
that the gravitational analogue of the electromagnetic spin
ineraction (\ref{Le}) is
\begin{equation}\label{gm}
{\cal L} _{gm}=\,\frac{\kappa}{8m}R_{abcd}S^{ab}S^{cd}\,.
\end{equation}
One can easily demonstrate that expressions (\ref{gm}) and
(\ref{agmr}) coincide (up to a factor $\,\ep/m\,$). It is
sufficient to this end to take into account the relation
$\,S^{ab}=\,\ep^{abcd}S_c u_d\,$, as well as the identity
\[ \tilde{R}_{abcd}=\,{1 \over 4}
\,\ep_{ab}^{\;\;\;ef}\ep_{cd}^{\;\;\;gh}
R_{efgh}=\,-\,R_{abcd}\, , \]
which is valid for a sourceless gravitational field.

It is natural to define in analogy with the magnetic moment
$$\frac{eg}{2m}S^{\mu\nu}\,,$$
the gravimagnetic moment
$$-\,\frac{\kappa}{2m}S^{ab}S^{cd}\,.$$
The gravimagnetic ratio $\kappa$, like the gyromagnetic ratio
$g$ in electrodynamics may have in general any value. Still, it is
natural that in gravity the value $\kappa=1$ is as singled out as
$g=2$ in electrodynamics. In any case, at $g=2$ and $\kappa=1$
the spin equations of motion have the simplest form.

For a classical object the values of both parameters $g$ and
$\kappa$ depend in general on its various properties. However,
for black holes the situation is different. It follows from the
analysis of the Kerr-Newman solution that the gyromagnetic ratio
of a charged rotating black hole is universal (and the same as
that of the electron!): $g=2$ \cite{cart}.

We will demonstrate that for the Kerr black hole the gravimagnetic
ratio is $\kappa=\,1$. This result follows in fact from the
analysis of the motion of spin of a black hole in an external
field carried out in Ref. \cite{hath} (though this statement by
itself was not explicitly formulated there). We will present here
an independent and, in our opinion, more simple derivation of this
important result.

We will consider the interaction (\ref{gm}) at large distance from
a Kerr hole at rest. To linear approximation in the created
gravitational field the corresponding Lagrangian density can be
written as
\begin{equation}\label{den}
\L=\,\frac{\kappa}{4m}[\s \times \nabla]_i
[\s \times \nabla]_j h_{ij} \delta(\r)\,.
\end{equation}
The thus induced correction to the energy-momentum tensor has
space components only:
\begin{equation}\label{ten}
\delta T_{ij}=\,\frac{\kappa}{2m}[\s \times \nabla]_i
[\s \times \nabla]_j \delta(\r)\,.
\end{equation}
Let us find the corresponding correction to the metrics. In the
gauge
\begin{equation}\label{cal}
\bar{h}^{\mu\nu},_{\nu}=\,0,\;\;
\bar{h}_{\mu\nu}=\,h_{\mu\nu}
-\,\frac{1}{2}\eta_{\mu\nu}h^\alpha_\alpha
\end{equation}
the static Einstein equations in the linear approximation are
$$\Delta \bar{h}_{\mu\nu}=\,16 \pi k T_{\mu\nu}\,.$$
The induced corrections to metrics equal
\beq\label{h00}
h_{00}=\,\kappa \frac{k}{m}(\s \nabla)^2 \frac{1}{r}\,,
\eeq
\beq\label{hmn}
h_{ij}=\,-\kappa \frac{k}{m}\left\{2\,[\s \times \nabla]_i
[\s \times \nabla]_j +\delta_{ij}(\s \nabla)^2\right\} \frac{1}{r}\,.
\eeq

Let us compare now the obtained corrections (\ref{h00}),
(\ref{hmn}) with the corresponding contribution to the Kerr
metrics. In the Boyer-Lindquist coordinates this metrics is
\begin{equation}\label{ker}
ds^2=\,(1-\,\frac{r_g r}{\Sigma})dt^2-\,\frac{\Sigma}{\Delta}dr^2-\,
\Sigma d\theta^2-\,(r^2+\,a^2+\,\frac{r_g ra^2}{\Sigma}\sin^2 \theta)
r^2 \sin^2 \theta+\,\frac{2r_g ra}{\Sigma}\sin^2 \theta d\phi dt\,,
\end{equation}
where $\Delta=\,r^2-\,r_g r+\,a^2\,,\;\;
\Sigma=\,r^2+\,a^2 \cos^2 \theta\,,\;\;\va=\,\s/m.$
At $r_g=\,0$ the metrics (\ref{ker}) describes a flat space in
spheroidal coordinates \cite{ll}. Meanwhile, it is Cartesian
coordinates which correspond in the flat space to the gauge
(\ref{cal}). The transition from the spheroidal coordinates to
Cartesian ones is carried out with the required accuracy by the
substitution
$$\r\rightarrow\,\r+\,\frac{\va(\va \r)-\,\r a^2}{2r^2}\,.$$
In the Cartesian coordinates the spin-dependent part of the
00-component of the metrics
$$ g_{00}=\,1-\,\frac{r_g}{r}+\frac{r_g a^2}{2r^3}\,
(3\cos^2 \theta-\,1)\,$$
coincides evidently with $h_{00}$ from formula (\ref{h00}) at
$\kappa =\,1$. For the space components of the Kerr metrics as
well their spin-dependent part reduces to the expression
(\ref{hmn}) at $\kappa =\,1$.

Let us note that the motion of the Kerr black hole in an external
gravitational field is not described by the Papapetrou equation
even if one leaves aside the problem of spin-orbit interaction
linear in spin. The point is that this equation refers to the case
$\kappa=\,0$ \cite{yb}.

It is proven in the same way that for a charged Kerr hole as well
the gravimagnetic ratio $\kappa=\,1$. Moreover, it can be proven
that the electric quadrupole moment of a charged Kerr hole equals
also that value,
\[ \,Q\,=\,-\,2\,{es^2 \over m^2}\,, \]
at which the interaction quadratic in spin decreases with energy.
One can demonstrate that other, higher multipoles of a
charged Kerr hole as well possess just those values which
guarantee for an interaction of any order in spin (but of course,
linear in an external field) the asymptotic decrease with energy.

\bigskip
\bigskip

We are grateful to I.A. Koop, R.A. Sen'kov, and Yu.M. Shatunov for
useful discussions. The work was supported by the Russian
Foundation for Basic Research through Grant No. 95-02-04436-Á.

\end{document}